\documentclass[aip,apl,reprint]{revtex4-1}

\usepackage{graphicx}
\usepackage{color}

\begin{document}


\title{Evidence for a capacitor network near the metal insulator transition in VO$_2$ thin films probed by in-plane impedance spectroscopy}

%


\author{J. -G. Ram\'irez}

\author{Edgar J. Pati\~no}
\affiliation{Departamento F\'isica, Universidad de los Andes, Bogot\'a D.C., Colombia}

\author{Rainer Schmidt}
\affiliation{Departamento F\'isica Aplicada III, Universidad Complutense de Madrid, 28040 Madrid, Spain}

\author{A. Sharoni}
\affiliation{Department of Physics, Bar-Ilan University, Ramat-Gan 52900, Israel}

\author{M. E. G\'omez}
\affiliation{Thin Film Group, Universidad del Valle A.A.25360, Cali, Colombia}

\author{Ivan K. Schuller}
\affiliation{Physics Department, University of California-San Diego, La Jolla California 92093-0319, USA}

\date{\today}


\newcommand{\Ce}{C$_E\,$}
\newcommand{\Wm}{\omega_\times}

\begin{abstract}
Impedance spectroscopy measurements were performed in high quality Vanadium dioxide (VO$_2$) thin films. This technique allows us investigate the resistive and capacitive contribution to the dielectric response near the metal-insulator transition (MIT). A non ideal RC behavior was found in our films from room temperature up to 334 K. A decrease of the total capacitance was found in this region, possibly due to interface effects. Above the MIT, the system behaves like a metal as expected, and a modified equivalent circuit is necessary to describe the impedance data adequately.  Around the MIT, an increase of the total capacitance is observed.
\end{abstract}


\maketitle


VO$_2$ exhibits a sharp first-order metal-insulator transition (MIT) and structural-phase transition (SPT) from a high temperature metallic rutile phase to a low temperature insulating monoclinic phase\cite{Masatoshi}. VO$_2$ also shows phase separation during the transition \cite{Qazilbash-science-2007}, which leads to percolation and avalanches \cite{sharoni:PRL:026404}. Potential applications range from bolometers \cite{Zerov:JEC:1999} to smart windows \cite{Babulanam1987347}. From the electronic point of view, the possibility to induce the MIT with an electric field \cite{Lee:APL2008}, opens a window in resistive nonvolatile random access memory (ReRAM) using VO$_2$ as a switching element \cite{Yu:nano:2009}. For applications in electronic devices the response of the MIT in VO$_2$ to an AC electrical field is crucial.  Impedance Spectroscopy (IS) enables the many contributions to the dielectric and resistive properties of electroceramic materials to be deconvoluted and characterized individually. 
The charge transport mechanism consists of several contributions such as sample-electrode interface and bulk resistance/capacitance. IS is shown to be a powerful tool to model such behavior. Recently,  impedance spectroscopy (IS) has been used to investigate the MIT in ZnO nano-powders \cite{nadeem:212104,Sahay:JMS:2008}, the resistive switching in NiO and TiO$_2$ thin films \cite{You:APL:2006,Lee:APL:2010}, magnetoimpedance in MgO tunnel barriers \cite{Snorri:APL:2010}, surface pasivation in silicon \cite{Kumar:apl:2010} and BiMnO$_3$ thin films \cite{Rainer:PRB:2009}. In each cases, IS gives relevant information about AC conduction.

Recently, AC voltage-current measurements were performed by Lee et. al. in two-terminal VO$_2$-based devices, where MIT oscillations were found \cite{Lee:APL2008}. They attributed these to be the result of the presence of non- VO$_2$ phases, which results in a large capacitive component around 156 pF.  A detailed study of the electrical response in a full-range frequency  in VO$_2$ thin films has not been done. In this letter a series of Impedance Spectroscopy (IS) measurements are described. These were performed in order to study the dielectric response of VO$_2$ thin films as a function of frequency of an AC electrical field across the MIT and deconvoluted several contributions.  

\begin{figure}[b]
	\centering
		\includegraphics[width=0.45\textwidth]{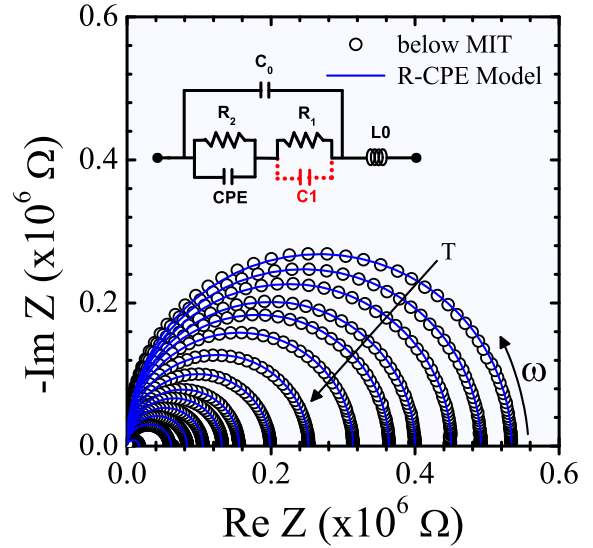}
	\caption{(Color online) Impedance Spectroscopy measurements of a 90 nm-thick VO$_2$ thin film at a frequency range 100 Hz - 1 MHz from 292 K to 338K (open circle). Line  (blue) corresponds to the R-CPE fit. The inset depicts the best equivalent circuit, C1 is indicated but not included in the fitting procedure}
	\label{fig:fig1}
\end{figure}
Highly (00$\ell$)-oriented vanadium di-oxide thin films were prepared by reactive RF magnetron sputtering from a 99.8\% vanadium target (1.5" diameter) on r-cut ($10\overline{1}2$) sapphire substrates. The films were deposited in a high-vacuum deposition system with a base pressure of 5$\times$10$^{-8}$ Torr \cite{ramirez:PRB:235110}. 

In-plane IS measurements were performed using an HP 4194A Impedance analyzer in the frequency range of 100 Hz to 10 MHz. The amplitude of the voltage signal was varied between 0.1 to 1 V; 0.5 V was found to be the optimal voltage in terms of a low signal-to-noise ratio without producing any significant heating to the sample. In order to avoid parasitic effects in the measurement a custom-built sample stage was used and 99.99\% pure Indium was used to connect 0.1-thick 10cm-long copper wires to the sample. Our sample stage allowed varying the temperature between 290 K and 360 K. 
Around 500 curves were taken in order to map the temperature dependence of impedance. IS data from VO$_2$ thin films between room temperature and 333 K are shown in Fig. \ref{fig:fig1} but only a few selected ones are shown. Once the temperature is increased the sample resistance starts to change and the impedance response shows smaller semicircles. 
Our data in Fig.1 shows signs of one single, slightly "non-ideal" dielectric relaxation, manifested by one slightly depressed semicircles for each temperature, where the center of each semi-circle has shifted below the x-axis (real-Z). In case of an ideal Debye-relaxation as represented by an ideal parallel resistor-capacitor element a perfect semicircle is expected \cite{Macdonald,Jonscher}.

In order to investigate the capacitance trend  as first approach we extracted the effective capacitance  \Ce  directly from the impedance data (see open-circle symbols in Fig.\ref{fig:fig1}) where VO$_2$ is in the insulating regime around the transition. The capacitance is obtained from the frequency  $\Wm$ at which the semicircles are maximum. In this region we have $\tau\Wm=1$ where $\tau=\mbox{R}\mbox{C}_E$. In this limit R corresponds to the real part of the impedance and \Ce is the equivalent capacitance plotted in Fig. \ref{fig:fig3}a. \Ce  slightly decreases between room temperature and 330 K and more interestingly increases appreciably as it approaches the MIT. Considering that only the sample and electrode interface are subject to temperature variations, any contribution to the impedance from the experimental setup (wires plus apparatus)  should  be temperature independent.  Therefore these observations reflect the capacitance changes in the VO$_2$ sample. 
\begin{figure}[tb]
	\centering
		\includegraphics[width=0.45\textwidth]{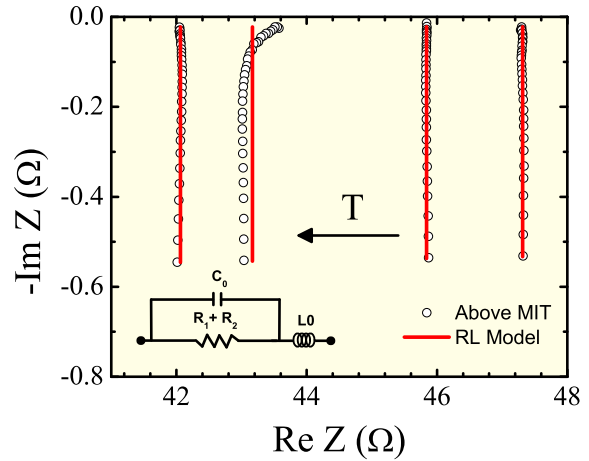}
	\caption{(Color online) High-temperature Impedance Spectroscopy measurements of a 90 nm-thick VO$_2$ thin film at a frequency range 100 Hz - 1 MHz from 335 K to 343K (open circle). Line  (red) corresponds to the RL model fit. Inset depicts the models elements.}
	\label{fig:fig1b}
\end{figure}
In order to further look into this we must deconvolute the different contributions to the data i.e. sample, interface and setup. With this aim an equivalent circuit was used to fit the experimental data. We used a different equivalent circuit in the VO$_2$ insulating and metallic phase. The best equivalent circuit for the IS data below the MIT is depicted in the inset of Fig. \ref{fig:fig1} (R-CPE model) \cite{RainerChap,Rainer:PRB:2009}. Here C0 corresponds to the stray capacitance of our setup, and was found to be around 0.1 pF at all frequencies. Note that given this value  it is very unlikely to observe the capacitance of the film at room temperature, because the in-plane configuration gives an estimated film capacitance close  to $5\times 10^{-17}$ Farads \cite{Note1}, too small to be detected by the instrument due to the stray capacitance of our setup. This value is assuming a single insulating phase material with a relative permittivity of 40. The R1-C1 element accounts for sample resistance-capacitance. C1 has been removed due to its small value compared to C0 (see circuit sketch in Fig. \ref{fig:fig1}). 

The Constant Phase Element (CPE) far below the transition has an impedance of the form $Z_{\mbox{CPE}} = (\mbox{C}_{\mbox{m}}(i\omega)^n)^{-1}$ where C$_{\mbox{m}}$ could be interpreted as Indium-VO$_2$ interface capacitance in  modified units [Fs$^{n-1}$]. R2 accounts for the interface resistance  and the exponent $n$ describes the "non-ideality" of the relaxation process  \cite{bosco:JEC:2008}. Above 334 K the R2 resistance  drops dramatically and the IS response is better described by an RL circuit as shown in Fig. \ref{fig:fig1b}. The imaginary part is now dominated by the wire inductance $L0$, around 1$\times$10$^{-4}$ H and the total resistance measured is a combination in series of the sample-bulk, interface and  wires. The model agrees satisfactorily to the experimental data. Four different curves are presented to show the change with temperature.
\begin{figure}[tb]
	\centering
		\includegraphics[width=0.45\textwidth]{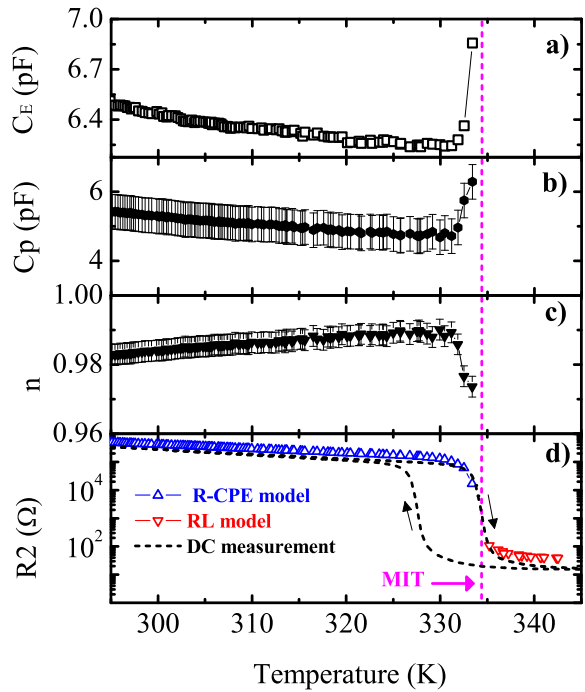}
	\caption{Temperature dependence of a) the effective capacitance  \Ce, b) the CPE capacitance C$_p$  in normal units, c) $n$ exponent, d) the resistance R1, extracted from R-CPE model (up triangle symbol) below the MIT and from the RL model (down triangle symbol) after the MIT. The dashed line correspond to a DC resistance measurement.  The vertical dashed-line indicates the MIT.}
	\label{fig:fig3}
\end{figure}

The fitting parameters extracted from the R-CPE model are shown in Fig. \ref{fig:fig3}b,c,d. Each point in the curves represents a fit at a specific temperature, the fitting-convergence criteria (chi$^2$) was kept as low as $10^{-4}$ for all points as shown in Fig. \ref{fig:fig1} (blue-solid line). C$_{\mbox{m}}$  was converted to C$_p$  in Farad units employing the relation $C_p =C_{\mbox{m}}(\omega_\times)^{n-1}$, for details see Ref. \onlinecite{units}. As can be seen from Fig. \ref{fig:fig3}b, the C$_p$ decreases monotonically between room temperature and 334 K. This decrease may be consistent with metallic domains nucleation at the interface and may be understood by considering a two dimensional interface as a network of mainly parallel capacitors. Metallic domain nucleation may then be understood as "switching-off" some of these insulating parallel capacitors and, consequently, the overall interface capacitance C$_p$ drops. By further increasing the temperature, C$_p$ increases appreciably. This behavior cannot be explained by interface effects and should be related to metallic domains nucleation at the bulk material. The insulating bulk may also be considered a capacitor-network and "switching off" some insulating capacitors leads to an increase of the overall sample capacitance, because such capacitors are now dominantly in series due to the in-plane measurement configuration. The film capacitance has been estimated though to be rather small, and, therefore, the increase in film capacitance needed to be massive (several orders of magnitude) in order to have a perceptible effect on the interface capacitance values shown ($\sim$ 6 pF) in Fig.\ref{fig:fig3}a,b. Further work using out-of plane measurements may be necessary to further investigate this. From \ref{fig:fig3}c it can be seen that the $n$ exponent slightly increases with temperature below 330 K below the MIT, but a sharp  decrease occurs when the temperature approaches 330 K; we believe that this is due to the high concentration of metallic domains which breaks the homogeneity across the sample. Fig. \ref{fig:fig3}d shows the temperature dependence of R2 in semilogaritmic scale. A monotonic decrease between room temperature and 334 K is followed by a sharp drop of almost two orders of magnitude within just 3 K. This is a clear manifestation of the MIT in our VO$_2$ films and in agreement with our previous work \cite{sharoni:PRL:026404}.  In order to confirm the R2 data from the fit, a  two-point DC measurement of resistance vs. temperature was done (dashed line Fig. \ref{fig:fig3}d). A percolation path is expected to appear somewhere in the middle of the MIT \cite{sharoni:PRL:026404} and once this path connects the electrodes, the resistance rapidly droops.

In conclusion, in-plane IS measurements were performed in high quality VO$_2$ thin films to successfully determine the dielectric response of the films at MIT. This technique allowed us to differentiate resistive and capacitive contributions to the electrical impedance of our VO$_2$ thin film samples and to investigate the dielectric response at the MIT near the percolation threshold. A slightly depressed semicircle behavior was found in our samples from room temperature up to 338 K. The decrease of the effective capacitance can be interpreted as an interface effect due to the metallic domains nucleation at the electrodes. Around 338 K metallic percolation across the system occurs leading to a resistance drop. Slightly below 338 K the curves of capacitance vs. temperature show an increase which could be related with formation of a network of capacitors across the electrodes. Slightly below 338 K  a  capacitor-network model  is able to reproduce the capacitance vs temperature  increase across the MIT. 

This work was supported by AFOSR, COLCIENCIAS, Excellence Center fot Novel Materials CENM. M.E. G\'omez wishes to acknowledge "El Patrimonio Aut\'onomo Fondo Nacional de Financiamiento para la Ciencia, la Tecnolog\'ia y la Innovaci\'on Francisco Jos\'e de Caldas'' Contract RC - No. 275-2011. R.S. wishes to acknowledge a Ramon y Cajal Fellowship from the Ministerio de Ciencia e Innovaci\'on in Spain.


%

\end{document}